\documentclass[twocolumn]{aastex63}

\usepackage{lipsum}
\usepackage{amssymb,amsmath,graphicx}
\usepackage{hyperref}
\usepackage{array}
\usepackage{footnote}

\makeatletter
\renewcommand*{\p@section}{\S\,}
\renewcommand*{\p@subsection}{\S\,}

\makeatother

\newcommand{\had}{{\sc Had}~}

\begin{document}

\title{
Multimessenger signals from black hole-neutron star mergers without significant tidal disruption
}

\author{William E. East}
\affiliation{Perimeter Institute, 31 Caroline Street, Waterloo, ON N2L 2Y5, Canada} 

\author{Luis Lehner}
\affiliation{Perimeter Institute, 31 Caroline Street, Waterloo, ON N2L 2Y5, Canada}

\author{Steven L. Liebling}
\affiliation{Long Island University, Brookville, New York 11548, USA}

\author{Carlos Palenzuela}
\affiliation{Departament  de  F\'{\i}sica $\&$ IAC3,  Universitat  de  les  Illes  Balears,  Palma  de Mallorca,  Baleares  E-07122,  Spain}


\begin{abstract}
We study the multimessenger signals from the merger of a black hole with a
magnetized neutron star using resistive magnetohydrodynamics simulations
coupled to full general relativity.  We focus on a case with a 5:1 mass ratio,
where only a small amount of the neutron star matter remains post-merger, but
we nevertheless find that significant electromagnetic radiation can be powered
by the interaction of the neutron star's magnetosphere with the black hole.  In
the lead-up to merger, strong twisting of magnetic field lines from the 
inspiral leads to plasmoid emission and results in a luminosity in excess of
that expected from unipolar induction.  We find that the strongest emission
occurs shortly after merger during a transitory period in which magnetic loops form
and escape the central region.  The remaining magnetic field collimates around
the spin axis of the remnant black hole before dissipating, an indication that,
in more favorable scenarios (higher black hole spin/lower mass ratio)
with larger accretion disks, a jet would form. 
\end{abstract}



\section{Introduction}\label{introduction}
The observations by gravitational wave (GW) detectors and conventional
electromagnetic telescopes of the binary neutron star (NS) merger known as GW170817 provided
a spectacular inauguration of the nascent enterprise of multimessenger astronomy~\citep{TheLIGOScientific:2017qsa,GBM:2017lvd}.
by constraining the equation of
state of high density matter, establishing that NS mergers source short gamma ray bursts,
and demonstrating the production of heavy elements from such mergers through 
observations of the subsequent kilonova. At the same time, nonvacuum binaries present
challenges to current GW observatories. In the case of binary NS systems, 
the lower total mass 
means that the stars come into contact at GW frequencies above a kilohertz,
where current detectors rapidly lose sensitivity. 

Black hole-neutron star (BHNS) binaries 
can potentially have much larger masses than binary NSs, and thus  
their merger is, in principle, easier to detect, as it takes place at lower frequencies where
current detectors are more sensitive~[$f\gtrsim 600 (10\ M_{\odot}/M_{T})$ Hz]. 
Importantly, for high enough binary mass ratios, the NS might be devoured by the black hole (BH) without being disrupted,
making the GW signal essentially indistinguishable from those sourced by a binary BH 
with the same masses.

For a sufficiently low mass ratio (depending on the BH spin and NS equation of state), 
the star is tidally disrupted outside the effective innermost stable orbit of the binary, resulting in 
the ejection of material from the system
and the formation of an accretion disk that may produce an electromagnetic counterpart, in particular
a short gamma ray burst. 
This scenario would be ideal for 
producing signals that would probe
the strong gravitational field (and arguably the strongest curvatures) as well as the 
properties of high density matter. 

However, observations of stellar mass BHs in binaries obtained to date (both in
electromagnetic observations of LMXBs and GW events by LIGO/Virgo) suggest a
dearth of such low mass ratio
systems~\citep{Abbott:2020gyp,2016A&A...587A..61C}. Instead, the evidence
suggests a prevalence of higher mass BHs ($\gtrsim6$ $M_{\odot}$) with low
spins. While this may be a result of observational bias, and the companions
in these binaries were not NSs (with one or two possible exceptions),
this suggests that disruption or other tidal effects might not
be strong enough to manifest in the GWs from most BHNS with high signal-to-noise, 
or to lead to an accretion powered transient.
For instance, the recent event GW190814~\citep{Abbott:2020khf}, with mass ratio of $\approx$10:1 had a companion
mass in the range of $\approx 2.50-2.67\ M_{\odot}$. The GW signal does not indicate whether the
companion was an NS or a BH, and its mass does not favor either one in particular unless further (potentially
biased) assumptions are made. Thus, this object was either the most massive NS, 
or the lightest BH, yet observed. Either answer would have profound consequences for astrophysics and nuclear physics.

The unknown nature of the secondary object in GW190814 demonstrates the importance of understanding potential electromagnetic
signals that could break such degeneracies.
Because the NS in a BHNS is likely magnetized~\citep{Spruit:2007bt}, its interaction
with a BH---which intensifies as the orbit tightens---could give rise to
electromagnetic counterparts.  Several authors have discussed how this might
occur within a unipolar induction~(UI)
model~\citep{Hansen:2000am,McWilliams:2011zi,Piro:2012rq,Lai:2012qe,DOrazio:2015jcb},
with the BH acting as a battery in a DC circuit with the NS.  However, whether
this simple steady-state picture is accurate remains an open question. In particular,
one interesting possibility that has not been explored in detail is that the
continuous twisting of magnetic field lines leads to more complicated
reconnection and plasmoid ejection, as well as emission channeled through the
development of a current sheet.  The twisting angle is related to the BH velocity
relative to the NS as $\zeta_\phi\approx 4 v_{\rm rel}/(\pi
c)$~\citep{Lai:2012qe}, and thus one expects these effects, and the departure
from the UI model, to be strongest around merger.

Several studies of the dynamics of magnetospheric interactions in binary NS
mergers have demonstrated how the binary's kinetic energy can be converted into
electromagnetic
radiation~\citep{Palenzuela:2013hu,Palenzuela:2013kra,Ponce:2014sza} through
UI and accelerating magnetic dipole
effects~\citep{Carrasco:2020sxg}, as well as how the twisting of magnetic flux
tubes can produce periodic flaring \citep{Most:2020ami}.

The BHNS case has been studied in the force-free approximation assuming a
helical Killing vector~\citep{Paschalidis:2013jsa,carrascoinprep}, 
which would approximately hold during the early inspiral.  Here, we
concentrate on the final, most dynamical stage of a 5:1 mass-ratio BHNS merger using full GR simulations.
This allows us to explore the
interaction of the BH with the NS's magnetosphere when it is strongest, as well as
the post-merger dissipation of the magnetosphere, which may
also source an electromagnetic 
transient~\citep{Lehner:2011aa,2011PhRvD..84h4019L,Pan:2019ulx}. We treat
the plasma with a resistive magnetohydrodynamics (MHD)
approach~\citep{Palenzuela:2012my} that can interpolate between the fluid
pressure dominated regions inside the NS, and the magnetically dominated regime
in the tenuous plasma surrounding the binary.

We find that the interaction of the BH
with the NS's magnetic field during merger 
leads to significant electromagnetic emission. 
The continual twisting of the magnetic field produces current
sheets with a complex configuration, occurring both in the vicinity of the BH
and also at larger distances. These current sheets, where
charges can be effectively accelerated, result as field
lines are stretched, forming X-points where reconnection takes place.
Plasmoids, isolated regions of closed field lines, are produced
from the reconnection, 
resulting in a level of electromagnetic emission stronger than that
estimated by UI.

\section{Setup}\label{details}
To study the system of interest, 
we implement the general relativistic,
resistive MHD equations, 
coupled to Einstein gravity. Thus, we capture the behavior of
both resistive, magnetized matter, and its interplay with the dynamical
spacetime. More details on the numerical methods and implementation are given in the appendix.
Unless otherwise stated, in the following we use Lorentz-Heaviside units with $G=c=1$.

We adopt initial data describing a nonspinning, quasi-circular binary consisting of a BH with
mass $M_\mathrm{BH}=7\ M_{\odot}$ and an NS with mass $M_\mathrm{NS}=1.33\ M_{\odot}$,
constructed with the Lorene library~\citep{lorene}.
The NS obeys 
a polytropic equation of state $p=K \rho^\Gamma$ with adiabatic index
$\Gamma=2$ and has a radius of $R_\mathrm{NS}=11.62$ km. 
The initial orbital frequency is $\Omega = 890$ Hz, and the binary undergoes roughly $2.5$ orbits
before merging.
Choosing a
mass ratio that allows a small amount of matter to remain bound but outside the
BH for some time enables us to study the development of magnetic
field structures post-merger that would be more marked, and relevant for
observations, in cases with lower mass ratios/higher BH spins.

The NS is given a dipole magnetic field with surface strength $B_*=3\times10^9$ G, though at this low value the magnetic
field does not have any significant effect on the spacetime~\citep{Ioka:2000yb},
nor on the hydrodynamics, except perhaps for the low density material in the vicinity
of the post-merger BH. 
Hence our results can be approximately scaled to arbitrary NS magnetic field values $B_9:=B_*/(10^9\ \rm{G})$ within this regime.

\section{Results}
The BHNS binary undergoes roughly two orbits before the outermost layer of the star
is stripped away and the bulk of the star is swallowed by the BH. This results in
a final BH with dimensionless spin $a/M \simeq 0.4$. Only a small
fraction of the matter remains outside the BH after merger ($\approx 5\% M_\mathrm{NS}$), and it is slowly accreting 
onto the BH at a rate $\dot M \propto
t^{-5/3}$ (there is negligible ejecta). 
While most of this behavior had been understood previously~\citep[e.g.][]{Chawla:2010sw}, our main focus here
is on examining the behavior of the electromagnetic field and the potential
electromagnetic signals induced by the merger. 

\begin{figure}
	\includegraphics[width=3.0in]{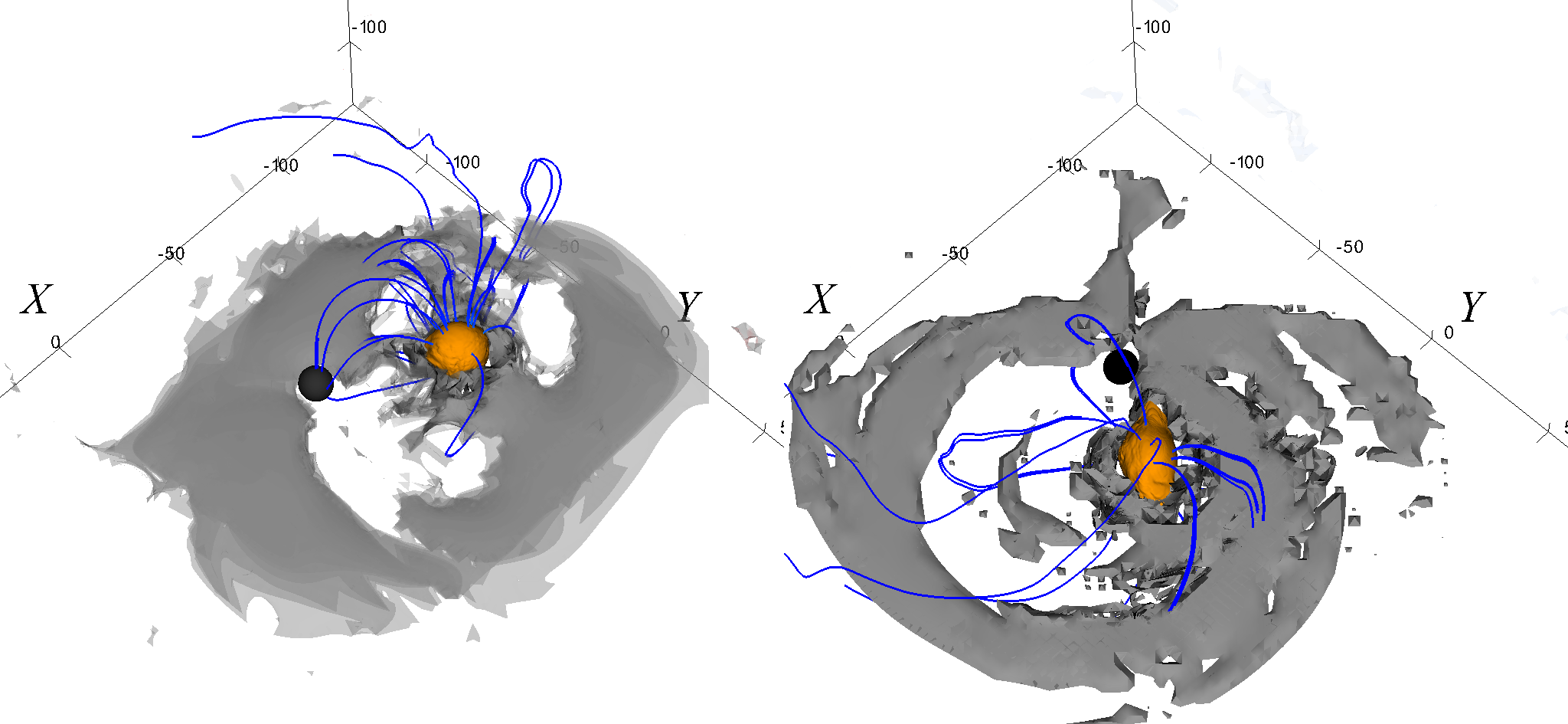}
	\caption{Snapshots at times $t=6.4$ (left) and $8.4$ ms (right). The figure illustrates the star (yellow, 
	isopycnic contour at $\rho= 3\times 10^{11}$ g/cm$^3$), the BH (black), magnetic field lines
	(blue, integrated over some seeds centered on the star), 
	and current sheet structure (defined by the norm of the current $J_i J^i$; gray, semitransparent). 
   The system orbits counterclockwise.
   The snapshots demonstrate the significant dynamics of the electromagnetic field, particularly the structure
   of the current sheets.
        A current sheet forms behind the 
	high curvature region of the BH, and trails the BH as it orbits. Another current sheet is correlated with the motion of the NS, but
	at some distance from the star. As the orbit proceeds, significant deformation of field lines leads to more current sheets.
	These sheets support closed magnetic field loops that arise from magnetic reconnection and that are transported away from the
	system.
	}
	\label{fig:currentsheets}
\end{figure}

Of particular relevance is the structure of current sheets associated with the
system, which not only bears a strong correlation with the dynamics of the
binary~\citep[as already indicated
in][]{Palenzuela:2013hu,Palenzuela:2013kra,Ponce:2014sza,Most:2020ami}, but
also with the characteristics of the compact objects involved. 
Figure~\ref{fig:currentsheets} illustrates this structure, showing current
sheets that have developed both on and off the orbital plane at two
representative times prior to merger. Roughly, we can understand the development
of these current sheets as due to two different effects.
On the one hand,
field lines emanating from the NS get sufficiently bent, as the star orbits, to
seed a current sheet some distance away, even though the NS is not spinning as
in previous binary NS studies. Analogous to the light-cylinder radius of an isolated
spinning NS, we expect this to occur at a lengthscale $L \simeq
\Omega^{-1}$ (where $\Omega$ is the angular velocity of the binary). 
On the other hand, field lines sufficiently
close to the BH get twisted to such a large degree as to seed a current sheet
in the wake of the BH's trajectory, even when the BH is not spinning~\citep[also previously observed in
simulations;][]{Palenzuela:2010nf,Neilsen:2010ax}.
As evident in Fig.~\ref{fig:currentsheets}, this latter current sheet is smaller
scale, comparable to the radius of the BH.

Because current sheets are the site of reconnection, their dynamics is key to
understanding the electromagnetic output of the system.  
As the orbit proceeds, field lines are more rapidly wound, and the magnetic
field strength in the strongly gravitating vicinity of the BH increases.  
In this low density region, our resistive scheme approaches the force-free limit.
With sufficient winding, X-point reconnection occurs and leads to closed field
loops that propagate away at near-luminal speeds, with loops forming near the
BH having greater field strength than those produced further
away\footnote{Greater field strength of loops formed near the BH has also been
observed in force-free solutions sourced by an effective NS
orbiting in a fixed Kerr BH spacetime~\citep{carrascoinprep}.}.
Previous studies of reconnection in the force-free approximation have found 
that the process is fast (relative, say, to near the ideal MHD limit),
with relativistic speeds roughly $v_\mathrm{rec} \approx 0.1$~\citep{Lehner:2011aa,Parfrey_2013}.
In particular, the reconnection speed is not set by the timescale of bulk dissipation
since in regions that are nearly force-free 
there is little Joule heating at the current sheets. 

\begin{figure}
	\includegraphics[width=3.0in]{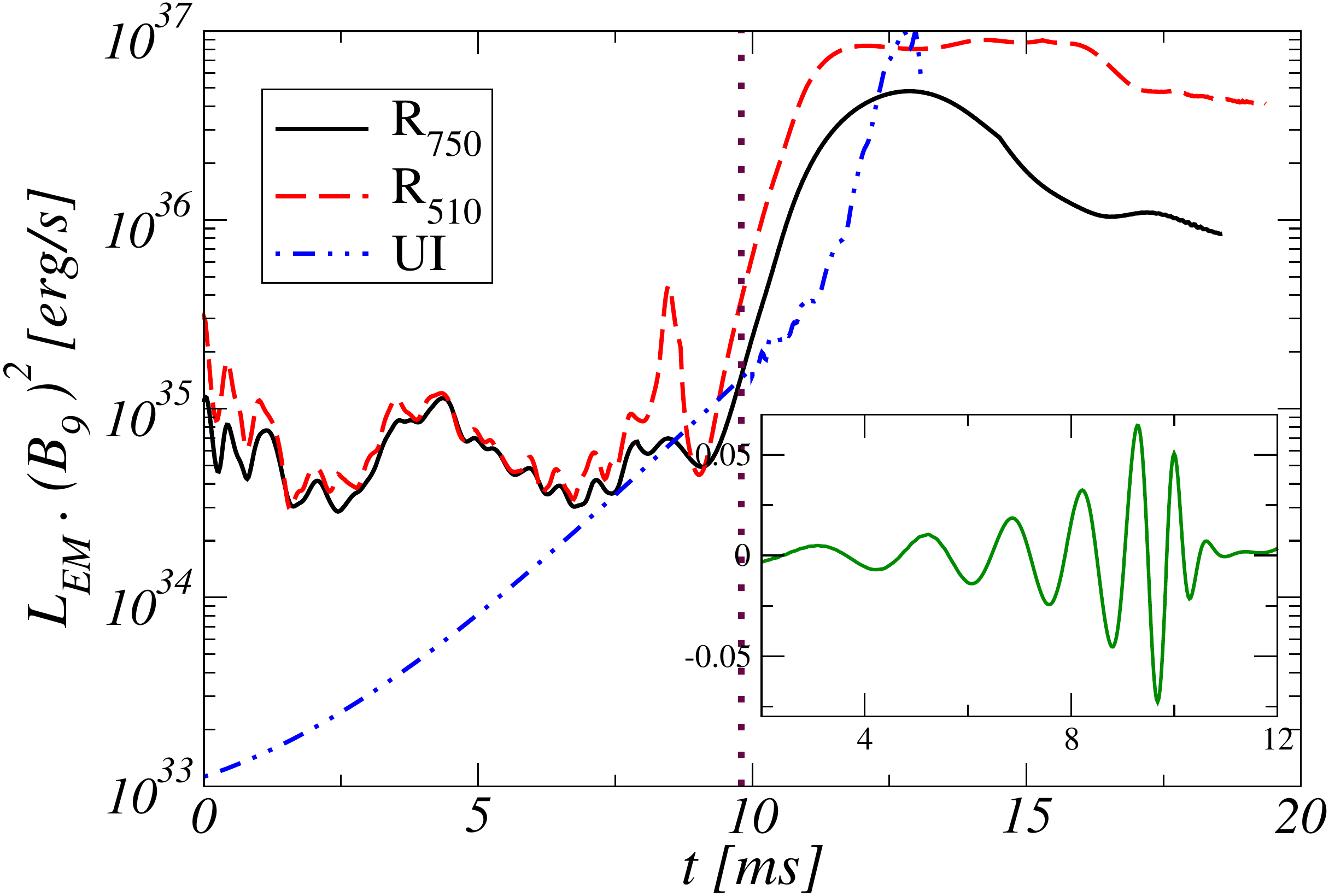}	
	\caption{
	Poynting flux luminosity measured at extraction surfaces with radii $R_e=510$~km and 750~km as
        a function of retarded time, along with the UI estimate (dashed)
	normalized to a surface field strength of $B=10^{9}$ G.
	The inset displays the plus polarization of the GWs (indicated by the real part 
	of the $l=m=2$ mode of $\Psi_4$). 
	The vertical line indicates the time at which the GW peaks. 
	Around this time, the electromagnetic luminosity
	grows steeply until it also reaches a peak. 
	}
	\label{fig:wavesGW_poynt}
\end{figure}

Examining the total Poynting flux from the system, shown at two different
spheres of observation in Fig.~\ref{fig:wavesGW_poynt}, we can see strong
modulations at several different frequencies in the lead-up to merger, which we
can attempt to associate with the abovementioned features of the current sheets.  
We expect high frequency features associated with the BH current sheet on
timescales of $\sim R_{\rm BH}/v_{\rm rec}\approx 0.5$ ms, while we expect lower
frequency modulations due to the more distant current sheets associated with NS 
orbit on timescales of $\sim \Omega^{-1}/v_{\rm rec}\approx 6$ ms 
(using $v_\mathrm{rec} \approx 0.1$ and $\Omega\approx1.6$ kHz). 
This is consistent with the variation in the electromagnetic flux 
seen in Fig.~\ref{fig:wavesGW_poynt}, though the latter is admittedly rather
noisy due to the complicated plasma dynamics.

The system radiates both in electromagnetic and gravitational channels, 
though the GW signal is dependent only on the acceleration of the source's quadrupole (and
not the magnetic field strength or reprocessing of electromagnetic outflows). The inset
of Fig.~\ref{fig:wavesGW_poynt} illustrates the plus polarization of the GWs.
The familiar chirp is present, and the post-merger, quasinormal ringing 
is consistent with the expectation for a remnant BH with $M\simeq8.3 M_{\odot}$ and $a/M\simeq0.4$~\citep[as predicted from simple arguments][]{Foucart:2012nc,Buonanno:2007sv}. 

Interestingly, the peak of the electromagnetic emission occurs a couple of milliseconds
later than that of the GWs. 
While the GW peak corresponds roughly to the maximum rate of change of the system's quadrupole
moment, the electromagnetic emission arises from the reconnection of magnetic field lines. Thus the
time delay of the electromagnetic emission is a consequence of the magnetic reconfiguration forced by
the formation of a common horizon for the BH and NS. We also note that the delay
is commensurate with the 
spin period of the remnant BH, roughly 2.5 ms.

 An important result of this paper is the comparison of this luminosity
with that predicted by the UI model $L_{\rm UI}$, which we also include
in Fig.~\ref{fig:wavesGW_poynt}.  Recall that the UI model prediction for
the binary studied here is $L_{\rm UI}\approx 4\times 10^{34} (v_{\rm
rel}/c)^2 B_9^2(100\ {\rm km}/d)^6$ erg/s.  We use the Keplerian
expressions for $\{v,r\}$ in terms of the orbital frequency $\Omega$.  We
estimate this latter one in terms of the GW frequency $f \approx \Omega
\pi$ (hence $L_{\rm UI}\propto f^{14/3}$), thus capturing the more rapid
rate at which $\{v,r\}$ change due to the increasingly strong
gravitational effects as the final plunge approaches.  The simulations
consistently show a larger luminosity than that of the UI model at
earlier times (lower frequencies), as that model does not capture the
complex phenomena associated with reconnections and the role of current
sheets.  Close to the coalescence, however, the two become nearly
equal\footnote{Note though that the assumptions underlying the UI
estimate become increasingly suspect at merger.}.  Furthermore, notice
that a slightly weaker Poynting flux is measured at larger extraction
spheres, reflective of the energy dissipated at current sheets located
between the spheres~\citep[a behavior recently pointed out
in][]{Most:2020ami}. Pre-merger, this integrated difference corresponds
to a dissipated energy of roughly $2\times10^{32}B_9^2$ erg (this
includes accounting for the energy stored in the region between the
extraction spheres, which is subdominant).
Post-merger, this difference is more significant, though this is likely,
at least in part, because the larger scale of the post-merger
electromagnetic field structures means that one must go to larger radii
to be in the wave zone and free of finite extraction radius effects.  In
the following discussion of the luminosity, we take the results from the
largest radius shown.

The significant post-merger emission evident in Fig.~\ref{fig:wavesGW_poynt} is
of the order of magnitude one would calculate for the Blandford-Znajek
luminosity~\citep{1977MNRAS.179..433B} using the initial magnetic field strength
of the NS. However, as discussed further below, the magnetic field in the
vicinity of the final BH is significantly lower, and this emission is actually
powered by the radiation of magnetic field loops. Previous studies of the
collapse of a magnetized NS in force-free or ideal MHD have found that the
resulting BH sheds its magnetosphere within a timescale $\approx 100M_{\rm BH} \approx 4.0$
ms$(M_{\rm BH}/8.3\ M_{\odot})$~\citep{Baumgarte:2002vu,Lehner:2011aa,2011PhRvD..84h4019L} or
less. This is of the same order, if somewhat shorter than, the timescale over
which the post-merger luminosity (at the largest extraction radius)
decreases. One factor contributing to this longer timescale compared to NS collapse
is the asymmetry of the merger scenario, which leads to magnetic field loops 
gathered to one side, instead of forming an axisymmetric, equatorial current sheet.

The luminosity is still significant at the end of the simulation, but
cannot be long lived if it is not powered by the BH and/or accretion. We can
obtain an upper bound on its persistance at this level using the
total magnetic energy stored in the NS's initial dipole as an estimate of the
available energy (ignoring that a significant amount is captured by
the BH) to obtain $U_{\rm dip}/L_{\rm EM}\approx100$~ms.  

\begin{figure}
	\includegraphics[width=3.0in]{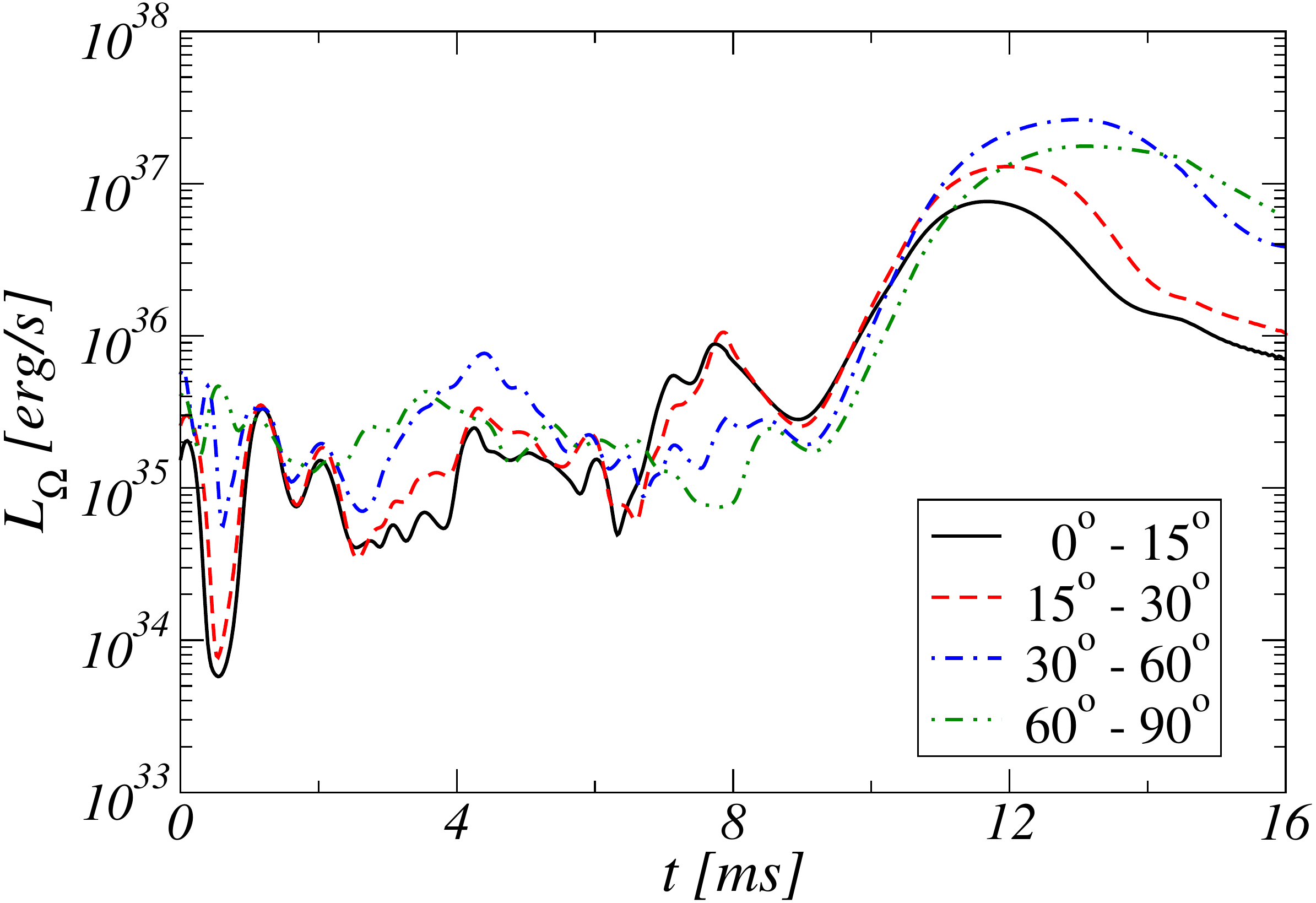}	
	\caption{Luminosity at an extraction radius of $R=750$ km through specified ranges of the polar angle (normalized by their angular size) 
        with $0^\circ$ corresponding to the direction of the total angular momentum of the system.
   During the inspiral phase, all regions show short timescale modulations, but the
   polar regions (in the direction transverse to the orbit) exhibit an overall increase that contrasts
   with the flatter level close to the equatorial (orbital) plane.
   At merger, the luminosity at all angles increases sharply,
   but subsequently decreases post-merger 
   as the BH accretes the little remaining matter, and the
	magnetic field lines are shed away or swallowed by the BH.
	}
	\label{fig:lumcones}
\end{figure}

To quantify the directional dependence, we compute the luminosity within specific
ranges of the polar angle, normalized by their angular size. 
In Fig.~\ref{fig:lumcones} we can see that there is an increase in luminosity with polar angle, indicating greater
emission near the orbital plane. The emission close to the orbital plane is also less variable
than the polar emission.
Near merger, the luminosity at all angles increases similarly rapidly.
Post-merger, the emission from
the equatorial region strongly dominates, and the luminosity at all angles then decreases
as the small amount of remaining
matter is accreted and magnetic field is shed or swallowed by the BH~\citep[see also][]{Lehner:2011aa}.

\begin{figure}
	\includegraphics[width=\columnwidth]{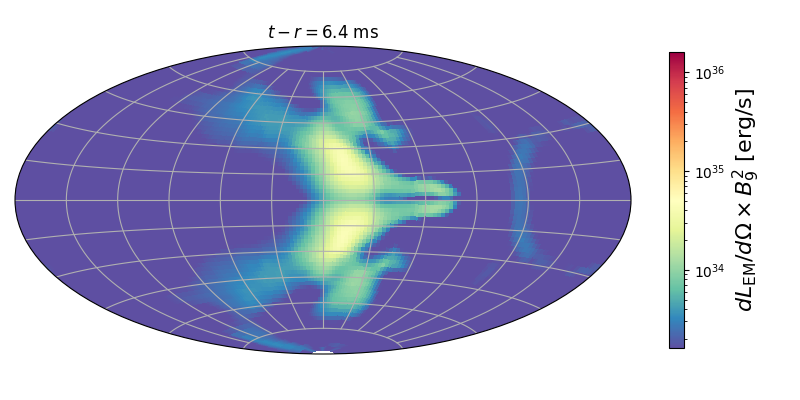}
	\includegraphics[width=\columnwidth]{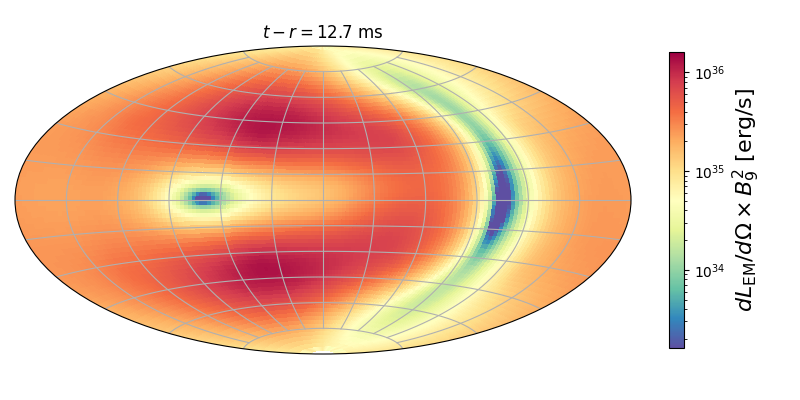}
	\includegraphics[width=\columnwidth]{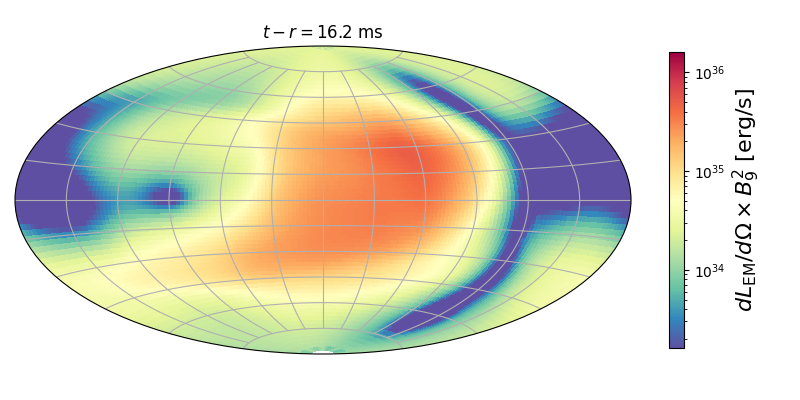}
	\includegraphics[width=\columnwidth]{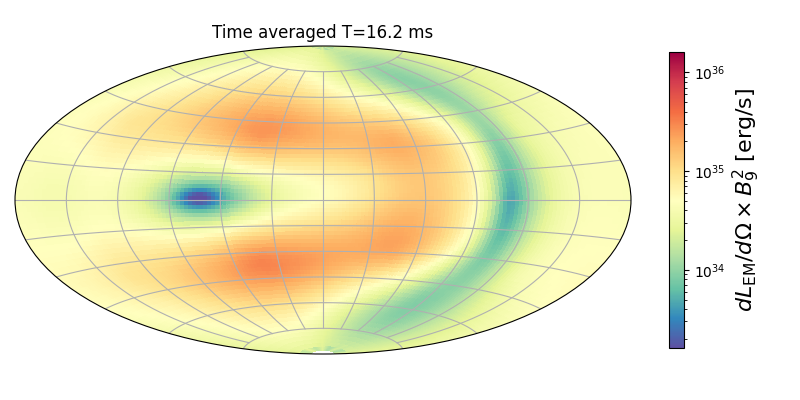}
	\caption{Snapshots of the Poynting flux at look-back times $t-r=6.4$, 12.7, and 16.2 ms (from top to bottom)
                 at an extraction sphere at radius $r=750$ km. The bottommost panel shows the time-averaged
                 flux for the duration $T=16.2$ ms, or equivalently the total radiated energy per steradian $\times1/T$.
	}
	\label{fig:sky_map}
\end{figure}

The angular dependence is further illustrated in Fig.~\ref{fig:sky_map}, which shows 
several snapshots of the angular dependence of the electromagnetic luminosity leading up to,
and following merger. In addition to the stronger emission in the vicinity of the equator, 
we can also see the strong nonaxisymmetric nature of the radiation.
However, from the time integrated flux in the
bottom panel of Fig.~\ref{fig:sky_map}, we can see that the emission is not strongly
observer dependent, in the sense that the energy emitted in most directions
is within an order of magnitude of the maximum. 

\begin{figure}
	\includegraphics[width=3.0in]{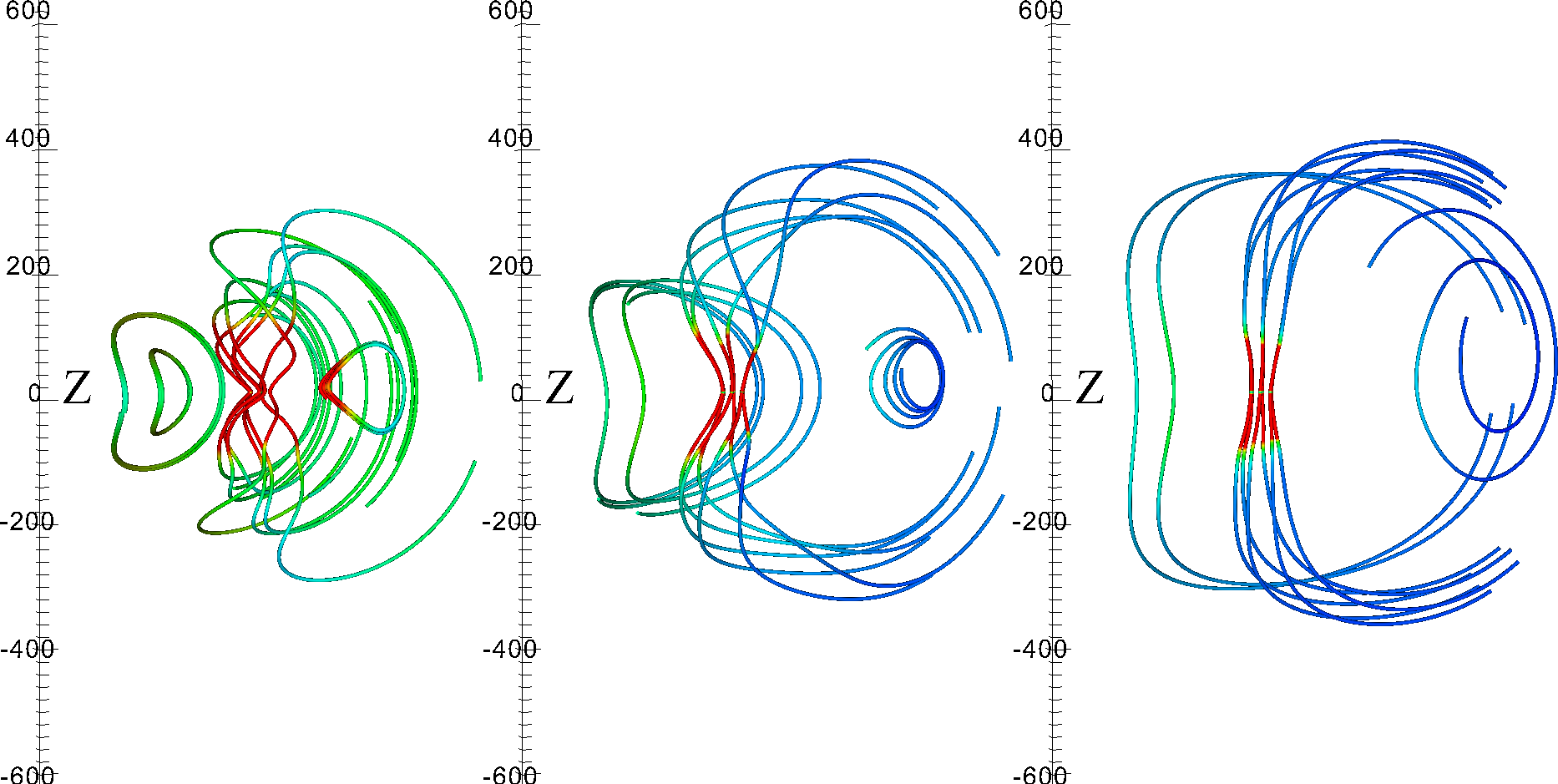}
	\caption{Snapshots of the magnetic field at times $t=13.3$, 16.4, and 19.0~ms from a distant vantage point. 
   A selection of magnetic field lines is shown, including lines passing near the BH (where the fields are the
   strongest)
   and a few additional lines away from the BH, which show the emission of a plasmoid.
   The color indicates the magnetic field strength ranging from $10^4B_9 $ G to $10^6B_9$ G, with
   all panels using the same colormap.
   The field topology in the vicinity of the BH gradually collimates along the
   BH spin axis.
	}
	\label{fig:twistingBH}
\end{figure}

Finally, it is interesting to examine the large-scale magnetic field post-merger, 
when there is a spinning BH with a small amount of matter remaining outside.
In Fig.~\ref{fig:twistingBH}, we display three snapshots of select magnetic field lines 
at times $\approx 3.5$, 6.5, and $9.1$~ms after merger (at intervals roughly corresponding to the
BH rotation period).
Most apparent, the magnetic field near the BH appears increasingly collimated and ordered, suggestive
of possible jet formation \`a la the Blandford-Znajek process.
However, this is only ephemeral for the system studied here
because of the dearth of matter to anchor the magnetic field. What little matter remains
has insufficient angular momentum and 
is being quickly accreted by the BH. With nothing to arrest 
the accretion~\citep[e.g.][]{10.1093/pasj/55.6.L69}, the
field lines from the BH become increasingly vertical with diminishing strength.
Instead of Blandford-Znajek emission, one can see the loops
formed by equatorial currents in the vicinity of the BH 
propagate away, still producing sizeable, though more transient, emission.
It would be interesting to study in more detail the near horizon plasma dynamics
giving rise to this emission, though we leave this for future study.

\section{Discussion}
This study focuses on a BHNS system sitting roughly at the boundary
between two regimes: one in which the NS fails to disrupt with essentially no electromagnetic
emission {\em powered by post-merger matter}, and the other in which the star fully disrupts leading to significant emission
due to the presence of an accretion disk and/or ejecta.
The latter scenario
is expected only for systems
with a sufficiently low mass ratio and/or high BH spin---a scenario so far seemingly disfavored by
GW observations (assuming the binary BH observations made so far
are representative of BHs in generic BHNSs). 
For such a disk, a standing question has been the degree and timescale
over which the magnetic field reorders and gives rise to a configuration favoring a jet~\citep[see, e.g.][
]{Paschalidis:2014qra,PhysRevD.98.123017,10.1093/mnras/stz2552,10.3389/fspas.2020.00046}.

In our case, where only a little material remains temporarily outside the BH, we
indeed see such a reordering.
Our evolutions suggest the formation post-merger
of poloidal structure which is a requirement of many models of BH jets~\citep[e.g.][]{Hawley:2015qma,Christie:2019lim}. 
Furthermore, we note that
in principle the amount of material in such a disk need not be too large. As argued in~\citet{Lee:2007js},
as little as $10^{-3}$ $M_\odot$ can anchor magnetic fields with strengths up to $10^{15}$~G. For the physical parameters studied here,
such an amount of matter, however, is accreted in a relatively short timescale. 
The fact that the magnetic field is successfully collimated even with such a small 
amount of material remaining is suggestive of what
would happen in more favorable configurations where
the NS is disrupted earlier and the amount of post-merger matter is larger.

 In the opposite direction (i.e. no spin and higher mass ratio), one would be left only with the type of electromagnetic counterparts
that would be induced by the features elucidated in this work: development of a complex current sheet, strong twisting of field lines, and
formation of X-points; all favoring the emission of plasmoids. As we point out, the energy associated
with such structures is enhanced due to the spacetime dynamics, thus improving the observational prospects.

We comment briefly on the potential for observing signals from the system. LIGO/Virgo is
able to observe GWs from such binaries at distances in excess of 100 Mpc, and this horizon will improve with further upgrades,
including dramatically with planned third generation detectors \citep[e.g.][]{Sathyaprakash:2019yqt}.
On the electromagnetic front, at high energies,
and taking the Burst Alert Telescope at Swift as an example,
its sensitivity to $100$~keV photons~\citep{Barthelmy:2005hs}
would allow for a detection up to a distance
of $D_L \simeq 50 (B_{12})$~Mpc\footnote{Skymaps of the potential high
energy radiation produced by the system will be presented elsewhere~\citep{ortizinprep}.}.
Such an estimate assumes perfect conversion efficiency; however, we note
realistic estimates are less optimistic by a factor of $\simeq 1\% -10\%$~\citep[e.g.][]{2019GCN.25341....1P}.

At lower frequencies, coherent radio emission has been suggested as a more
likely observational prospect. As discussed
in~\citet{Most:2020ami} and~\citet{Sridhar:2020uez}, both magnetic reconnection in the
current sheet and the synchrotron maser process due to plasmoids shocking the
ambient plasma are potential mechanisms for such emission. 

\bigskip

\begin{acknowledgments}
We thank Federico Carrasco and Alexander Philippov for discussions. LL thanks the CCA at the Flatiron
Institute for hospitality during an early stage of this work.
SLL is supported by the NSF under grants  PHY-1912769
and PHY-2011383. 
CP acknowledges support from the Spanish Ministry of Economy and Competitiveness grants AYA2016-80289-P
and PID2019-110301GB-I00 (AEI/FEDER, UE). 
WE and LL are supported in part by NSERC through a Discovery Grant, and LL also thanks CIFAR for support.  
Computations were performed on XSEDE resources and the Niagara supercomputer at SciNet.   
SciNet is funded by the Canada Foundation for Innovation; the Government of Ontario; Ontario Research Fund - Research Excellence; and the University of Toronto.
Research at Perimeter Institute is supported by the Government of Canada and by the Province of Ontario
through the Ministry of Research, Innovation and Science.

\end{acknowledgments}

\appendix
In this section, we give more details on our scheme for evolving a binary
with a BH and magnetized NS, including the numerical implemention.
We employ the \had computational
infrastructure~\citep{had_webpage} and 
implement the general relativistic,
resistive MHD equations, as described
in~\citet{Palenzuela:2012my} and~\citet{Palenzuela:2008sf}, coupled to Einstein gravity in
the CCZ4 formulation~\citep{alic12,bezares17}. 
For a more thorough description of the implementation we refer the reader to
aforementioned references.

The magnetized star is described by the total stress-energy tensor
\begin{eqnarray}\label{stress-energy-perfectfluid}
T_{\mu \nu} &=& \left[ \rho (1 + \epsilon) + p \right]
u_{\mu} u_{\nu} + p g_{\mu \nu}
+{F_{\mu}}^{\lambda} F_{\nu \lambda}
- \frac{1}{4} g_{\mu \nu} ~ F^{\lambda \alpha} F_{\lambda \alpha} 
\end{eqnarray}
where $F^{\mu \nu}$ is the Faraday tensor,
which can be decomposed in terms of the
electric $E^{\mu}$ and magnetic $B^{\mu}$ fields\footnote{Notice that a factor of $1/\sqrt{4\pi}$ has been absorbed in the definition 
of the electromagnetic fields.}. Here $u^a$ is the fluid four velocity, $\rho$ is
the rest mass density, $\epsilon$ the internal energy, and $p$ is the pressure.
Here we use a $\Gamma=2$ equation of state $p=\rho \epsilon$. 

The evolution of the magnetized matter must obey both the Maxwell equations and the conservation of total stress-energy tensor. 
Going beyond the ideal MHD limit, which treats the fluid as a perfect conductor, requires
a prescription for the electric current to close the system of equations, called resistive MHD.
The ideal MHD and the force-free limits can be captured with the
phenomenological current introduced in~\citet{Palenzuela:2012my}, which includes the isotropic conductivity and (some of) 
the anisotropic Hall terms
\begin{equation}\label{ohm_relativistic_spatial}
J_i =  q \left[(1-H) v_i + H v^{d}_i  \right]
+ \frac{\sigma}{1 + \zeta^2}
\left[{\cal E}_i + \frac{\zeta^2}{B^2} (E^k B_k) B_i\right]
\end{equation}
where $v^d_i= \epsilon_{i j k} E^j B^k /B^2$ is the drift velocity, $v_i=u_i/W$ is the Eulerian velocity with
associated Lorentz factor $W=\alpha u^t$ (in terms
of the lapse $\alpha$), and we have introduced the shorthand 
\begin{eqnarray}\label{newEB}
{\cal E}_i &=& W  \left[ E_i + \epsilon_{i j k} v^{j} B^{k}
- (v_{k} E^{k}) v_{i} \right] ~.~~~~ 
\end{eqnarray}
The kernel function $H$ is defined such that it smoothly varies with density
from zero inside the star, to unity outside, with a very high isotropic
conductivity $\sigma =2\times 10^{10}\ {\rm s}^{-1}$ and an anisotropic ratio $\zeta =
H \sigma \tau$, where we set $\tau=10^5\sigma^{-1}=5\times 10^{-6} \ {\rm s}$ to be 
shorter than the dynamical time of the binary, but much longer than $\sigma^{-1}$. 
In our particular scenario, in the interior of the star (i.e., $H\approx 0$), the large isotropic conductivity effectively reduces the system to the ideal MHD limit. 
In the exterior of the star (i.e., $H\approx 1$) the anisotropic terms dominate and effectively enforce the force-free condition.

For the initial magnetic field, we adopt a magnetic moment $\bf
\mu$ that describes a dipolar magnetic field $\bf B$ in the comoving frame of
the star. This magnetic moment is aligned with the orbital angular momentum.
The magnitude of the dipole moment is related to the
radial magnetic field at the pole of the star $B_*$, by the relation $\mu= B_*
R_\mathrm{NS}^3$. 
The electric field is obtained from the ideal MHD condition $\bf E=-
\bf v \times \bf B$, where the velocity in the star is given by the orbital
motion, and we assume that the magnetosphere is initially at rest.
In our simulations, we take $B_*=3\times10^9$ G though, as mentioned above, our results 
can be approximately scaled to arbitrary NS magnetic field values.

The gravitational equations are discretized
with fourth-order accurate finite difference operators, while high-resolution shock capturing  methods  based  on  the  HLL  flux  formula  with  PPM  reconstruction are used to discretize the resistive MHD equations~\citep{Palenzuela:2012my}. 
The time evolution is performed through the method of lines using a third-order accurate Implicit-Explicit Runge-Kutta integration scheme~\citep{ParRus:2005} in order to deal with the stiffness of the resistive equations~\citep{Palenzuela:2008sf}. In our production run, 
the adaptive mesh refinement criteria tolerance is chosen to guarantee that the star is covered by 84 points in each 
direction. The computational domain $[-1200,1200]^3 \mathrm{km}^3$ is discretized with seven refinement levels (in 
addition to the coarsest grid, which covers the whole domain) with a 2:1 refinement 
ratio. The coarsest grid has a grid spacing of $\Delta x = 30$~km and the finest grid (which covers the compact objects)
has a grid spacing of $\Delta x \approx 0.23$~km.
We adopt a Courant parameter $\Delta t/\Delta x=0.25$ in each refinement  level.
To check the consistency of our results, we also evolve (i)~the same computational set-up with one less refinement level for the full inspiral and merger; and
(ii)~another with seven refinement levels, but in which the computational domain extends only to $\pm600$\,km and the coarsest resolution is $\Delta x = 20$~km, for roughly the first orbit.
The results of these additional simulations suggest that our production run is in the convergent regime.

\bibliographystyle{aasjournal}
\bibliography{references}

\end{document}